# Self frequency-locking of a chain of oscillators

S. Giurdanella and A.C. Sparavigna
Dipartimento di Fisica, Politecnico di Torino
C.so Duca degli Abruzzi 24, Torino, Italy

**Abstract**
The paper studies the vibrational modes of a slightly damped uniform chain, with $n$ masses coupled by elastic forces. It will be shown that, for certain lengths of the chain, that is for certain values of $n$, the damping of one of the masses at a specific position in the chain is able to constrain the vibration of the system to oscillate at a specific frequency. The damped mass turns out to be a node of the chain, subdividing it in two parts. This node can be considered as the synchronization element of the two subchains. As a consequence the oscillating system of $n$-masses is self-locking to the synchronized frequency of its subchains.

The dynamics of chains with harmonic oscillators is intuitive and quite easy to solve. For this reason, these chains remain the preferred models to use in solid state physics to describe the vibrational properties of crystalline lattices and their quantization in phonons. Modelling with harmonic chains is also a rather old research approach. As reported by Brillouin, harmonic chains were the subject of Isaac Newton's researches on the speed of sound [1,2]. Anharmonic models instead are not so simple or so old: it was about 1954, that Enrico Fermi proposed to study numerically a chain of particles linked by springs with weak nonlinearity. The so-called Fermi-Pasta-Ulam-Tsingou problem marked the beginning of studies on nonlinear physics, solitons and chaos [3-5].

Besides modelling the phononic systems, we can assume the harmonic chains as suitable models for metamaterials, that is, for those artificial compounds engineered to gain their properties from the structure rather than from the composition. Most of metamaterials are obtained from a periodic arrangement of few components. If we consider the one-dimensional case for instance, we immediately recognize that a chain of masses and springs is the simplest metamaterial we can imagine. In fact the dispersion of frequencies is coming from the structure of the chain, not only from the stiffness of the springs. Moreover, the dynamics governing a mechanical chain with masses, springs and dampers, can be immediately transferred to modelling the transmission lines (see Fig.1), built by means of electric circuits, with lumped capacitive and inductive components [6].

For all these reasons, harmonic chains are still deserving new investigations. In this paper, we propose a discussion on the vibrational modes of chains, starting from the results shown in Ref. [2]. The data reported in that paper are confined to a low number of masses. The model of the chain was a uniform one, made up of $n$ equal $m$ masses and $n$ equal linear springs with stiffness $k$ (see Fig.1, upper part). A damping contribution was attached, as in Fig.1, to a single mass of the chain. Solving the dynamical matrix of the system, the complex eigenvalues corresponding to the damped oscillations can be determined. The real part of the eigenvalue gives the life-time of oscillation, the imaginary part its proper angular frequency.

In Ref.2, the number of masses was small ($n$=5). We extended the calculation to an arbitrary number of masses, observing that some undamped oscillating modes survive in chains with certain lengths, when the damping contribution is attached to a mass at a specific position. The consequence is that the frequency of the chain is self-locking to specific values. These values are determined by the length of the chain and the position of the damper. That is, we will see that, for chains with a given length, the damping of one of the masses at some specific positions in the chain constrains the vibration of the system to self-lock at specific frequencies. Moreover, the damped mass turns out to be a node of the chain. For this reason, the chain can be viewed as subdivided in two parts, that is, as if it were composed by two connected oscillating chains, which are synchronized to the same frequency. The oscillating system is self-locking to the synchronized frequency of its subchains. The damped mass is the synchornization element. Instead of synchronizing groups of coupled oscillators with sparse connections, as recently proposed [7], the synchronization can be achieved by means of a dissipative element in the chain. Since transmission lines are described by the same equations of mechanical chains, the presently discussed model could be proposed for their synchronization.

Models with dampers can be interesting for other engineering applications. In civil engineering for instance, a damper is an effective control device commonly attached to a vibrating primary system for suppressing undesirable vibrations induced by winds and earthquake loads [8-10]. The study of nodes and of the self-locking at specific frequencies in vibrating models is fundamental for the quality of many civil structures.

The model from Ref.2 is the mechanical one in Fig.1. Masses are on a horizontal frictionless surface. The chain starts with a mass connected to a wall and ends with a mass, connected only with the previous one. A damping friction can be added to each mass. The authors studied the case with a single damped mass and with all the masses damped, giving all the eigenvalues for chains till $n=5$. The aim of that paper was the proposal of an approximate analytic estimation for eigenvalues.

Tab.2 in Ref.2 is showing the values in the case with $n=4$ and $p$ ranging from 1 to 4. In the case that the damping effect is placed on the third mass, one of the complex conjugated pair of eigenvalues is purely imaginary. This means that the corresponding mode survives on long times, when all the others modes have been damped by their real part different from zero. The authors did not discuss the presence of this undamped mode and the reason for its existence. We decided then to investigate the presence of pure oscillating modes for longer chains, in the case of a single damped mass (the case with all the masses damped have no undamped modes).

We use the same equations as proposed in Ref.2 (here reported for the reader convenience in the Appendix): first of all we searched existing undamped modes when the number $n$ of masses increases. The $n$ conjugated pairs $\lambda = d \pm iw$ eigenvalues were obtained in two ways, by means of a MATLAB® program and with the QR algorithm for real Hessenberg matrices as in Ref.11, for Fortran programming. For each $n$, we supposed a single damped mass with its position $p$ spanning from 1 to $n$. We found that for $n=4, 7, 10, 12, 13, 16, 17, 19, 22, 24, 25, 27, 28$ .... these modes exist, that is, there are purely imaginary eigenvalues ($d=0$). Fig.2 reports on the left, the behaviour of the real (x-axis) and of the imaginary (y-axis) of the eigenvalues, for the corresponding choices of $n$ and $p$ as reported in the figure, and for the specified physical parameters. Note that the real part of the eigenvalues decreases in magnitude when the number of masses increases, this is due to the fact that we have damped only one mass in the chain and its effect is reduced increasing its length.

The right part of the figure shows the behaviour for a specific chain with $n=19$ and $p=18$, which has eigenvalues with a null real part, and that of the two following chains $n=20$ and $n=21$, which do not possess undamped modes. There are eigenvalues $\lambda = d \pm iw$ having a quite small real part, $d$, definitely different from zero. We see modes with a null real part also for other positions of the damped mass, different from $p=n-1$, as reported in Table I for chains with few masses.

Let us start discussing more deeply the case of a chain with 4 masses ($m_1, m_2, m_3, m_4$), the third ($m_3$) damped. Fig.3.a reports the behaviour of the third ($m_3$) and fourth masses ($m_4$) as a function of time, for $m=1$ kg, $k=1$ N/m, $c=0.01$ s$^{-1}$. The position $x_3$ is damped to zero. Image 3.b reports the behaviour of the positions ($x_2, x_4$) of the oscillating masses ($m_2, m_4$), with the time as parameter. The initial point of the motion is at the right upper corner marked by a red dot, whereas the limit of the motion for asymptotic times is described by the grey diagonal line. Fig.3.c and 3.d report the phase space diagrams for masses $m_4$ and $m_3$ respectively, where 3.d has a clear limit-cycle, which is observed also for $m_2$. In the case of 3.d, the limit is the central point of the image, coherently with the behaviour in 3.a.

The evaluation of the corresponding dynamical matrix is easily to make: the pure imaginary eigenvalue has magnitude $w=\sqrt{(k/m)}$. The corresponding eigenvector is parallel to (1,1,0,-1). This means that we can subdivide the chain and the corresponding dynamical problems in two parts, one concerning an oscillating system composed by the first two masses ($m_1,m_2$) and the other by the last mass $m_4$, as exemplified in Fig.4. The eigenvalues of the first oscillator have magnitudes $w=1.000$ and 1.732, the second has $w=1.000$. It is then possible for the two parts (A and B) of the chain to oscillate with the same frequency, $w=1.000$, and with a phase suitable to let mass 3 at its equilibrium position. That is, $m_3$ becomes a node for the longitudinal oscillations of the chain.

As we can see from Table I, the same behaviour is observed for longer chains. In this case there is the possibility to have part B composed by more than one single mass: this fact increases the number of possible frequencies. Table II displays all the modes of the two sub-chains A and B. The common values are marked with the bold character. Fig.5 is visualising the same result as in Table II, with the comparison of dispersion of the chain without dampers and dispersions of the two sub-chains. The figure and the table show the synchronized frequencies. From these results, it is clear that the synchronization of the two parts of the chain is possible only for specific values of $n$ and $p$. The following Fig.6 is reporting the undamped modes appearing in the case of very long chains. From such charts, one has the possibility to decide where a damper must be placed to have the oscillation of the system to self-lock at one of the synchronization frequency.

Let us conclude remarking that we are dealing with a free-end chain: this mechanical system has its analogous in a transmission line with the open-end termination. It is possible then that the behaviour displayed by our model could have interesting for applications in devices including transmission lines, such as for many mechanical structures.

**Appendix**

The dynamical equations for a linear, discrete mechanical system with n degrees of freedom is given by

$$\mathbf{M}\ddot{\mathbf{q}}(t) + \mathbf{C}\dot{\mathbf{q}}(t) + \mathbf{K}\mathbf{q}(t) = 0 \tag{A1}$$

where **M**, **C** and **K** are $n \times n$ mass-, damping- and stiffness matrices, respectively, and **q**(t) is the n-dimensional coordinate vector. Let us assume

$$\mathbf{q} = \tilde{\mathbf{q}}e^{\lambda t} \tag{A2}$$

leads to the eigenvalue problem

$$[\lambda^2 \mathbf{M} + \lambda \mathbf{C} + \mathbf{K}\tilde{\mathbf{q}}] = 0 \tag{A3}$$

Where $\lambda$ and $\tilde{\mathbf{q}}$ denote an eigenvalue and the corresponding eigenvector, respectively. The case that we discussed in this paper has a viscous damper at the $p$-th mass. The motion of the oscillator is governed by Eq.A1 with matrices in the form:

$$\mathbf{M} = \text{diag}(m), \quad \mathbf{C} = \text{diag}(\delta_{ip}c), \quad \mathbf{K} = k\begin{bmatrix} 2 & -1 & & & 0 \\ -1 & 2 & -1 & & \\ & \ldots & \ldots & \ldots & \\ & & \ldots & 2 & -1 \\ 0 & & & -1 & 1 \end{bmatrix}, \qquad (A4)$$

We follow as in Ref.2 the eigenvalue problem formulation from Müller and Schiehlen, Shabana and Newland (see Ref.12 for details). The state vector y($t$) be defined as the ($2n \times 1$) vector:

$$\mathbf{y} = \begin{bmatrix} \mathbf{q} \\ \dot{\mathbf{q}} \end{bmatrix} \qquad (A5)$$

The ($2n \times 2n$) system matrix is given as:

$$\mathbf{A} = \begin{bmatrix} \mathbf{0} & \mathbf{I} \\ -\mathbf{M}^{-1}\mathbf{K} & -\mathbf{M}^{-1}\mathbf{C} \end{bmatrix} \qquad (A6)$$

The physical parameters are $m$, $k$, and $c$, damping coefficient. A MATLAB® script has been implemented to solve numerically the corresponding eigenvalues and eigenvectors problem. The dimension of matrix, the position $p$ of damping, mass, stiffness and damping coefficient were the parameters to use in the numerical program. At the end of calculation, real and imaginary parts of eigenvalues were plotted by means of a dedicated MATLAB® tool, or a tabular vision was considered. A Fortran programming was also used, with the QR algorithm for real Hessenberg matrices [11], to increase the speed of calculations.

**References**
1. L. Brillouin, Wave Propagation in Periodic Structures, 2nd Edn. Dover, New York, 1953.
2. M. Gürgöze and A. Özer, On the slightly damped uniform n-mass oscillator, Computers & Structures, Vol.5, pp.797-803, 1996.
3. T. Dauxois, Fermi, Pasta, Ulam and a mysterious lady. Physics Today, Vol.61, pp.55-57, 2008.
4. J. Ford, The Fermi-Pasta-Ulam problem: Paradox turns discovery, Phys. Rep., Vol.213, pp.271-310, 1992.
5. T.P. Weissert, The Genesis of Simulation in Dynamics: Pursuing the Fermi-Pasta-Ulam Problem, Springer, 1997.
6. E. Weber and F. Nebeker, The Evolution of Electrical Engineering, IEEE Press, Piscataway, New Jersey USA, 1994.
7. Z. Zheng, X. Feng, B. Ao and M.C. Cross, Synchronization of groups of coupled oscillators with sparse connections, Europhysics Letters, Vol.87, pp.50006-1-50006-5, 2009.
8. H. Frahm, Device for damping vibration of bodies, U.S. Patent No.989-958, 1911.
9. R.J. McNamara, Tuned mass dampers for buildings, Journal of the Structure Division, ASCE, Vol.103 pp.1985–98, 1977.
10. P.D. Cha and C. Pierre, Imposing nodes to the normal modes of a linear elastic structure, Journal of Sound and Vibration, Vol.219, pp. 669-687, 1999.
11. Numerical Recipes in Fortran 77, Second Edition, 1992, http://www.nrbook.com/a/bookfpdf.php.

12. M. Gürgöze, On various eigenvalue problem formulations for viscously damped linear mechanical systems International, Journal of Mechanical Engineering Education, Vol.33, pp.235-243, 2005.

**Tables**

| Number of masses, *n* | Damped mass, *p* | Eigenvalues (*Im*) *w* |
|---|---|---|
| 4 | 3 | 1.000 |
| 7 | 3 | 1.000 |
| 7 | 5 | 0.618, 1.618 |
| 7 | 6 | 1.000 |
| 10 | 3 | 1.000 |
| 10. | 6 | 1.000 |
| 10 | 7 | 0.445, 1.247, 1.802 |
| 10 | 9 | 1.000 |
| 12 | 5 | 0.618, 1.618 |
| 12 | 10 | 0.618, 1.618 |
| 13 | 3 | 1.000 |
| 13 | 6 | 1.000 |
| 13 | 9 | 0.347, 1.000, 1.532, 1.879 |
| 13 | 12 | 1.000 |

Table I: Values of the undamped mode eigenvalues in the case of *m*=1 kg, *k*=1 N/m, *c*=0.01 s$^{-1}$.

| | |
|---|---|
| *n*=7, *p*=5<br>*w*: 0.618, 1.618 | part A (4 masses), *w*: **0.618**, 1.176, **1.618**, 1.902<br>part B (2 masses), *w*: **0.618**, **1.618** |
| *n*=10, *p*=7<br>*w*: 0.445, 1.247, 1.802 | part A (6 masses), *w*: **0.445**, 0.868, **1.247**, 1.563, **1.802**, 1.950<br>part B (3 masses), *w*: **0.445**, **1.247**, **1.802** |
| *n*=13, *p*= 9<br>*w*: 0.347, 1.000, 1.532, 1.879 | part A (8 masses), *w*: **0.347**, 0.684, **1.000**, 1.285, **1.532**, 1.732, **1.879**, 1.969<br>part B (4 masses), *w*: **0.347**, **1.000**, **1.532**, **1.879** |

Table II: Values of the undamped mode eigenvalues in the case of *m*=1 kg, *k*=1 N/m, *c*=0.01 s$^{-1}$, for three chains with the specified position of the damped mass and the two corresponding parts A and B of them. The synchronized frequencies are marked by the bold character.

**Figures**

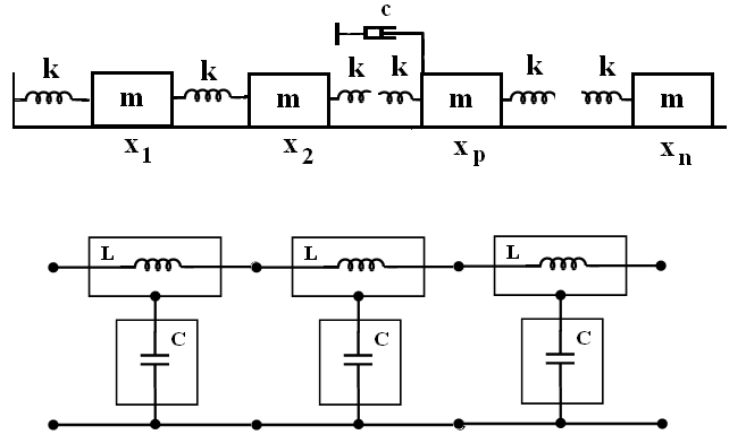

Fig.1: Chain of coupled oscillators with a single damped mass, in the upper part, in the upper part of the image. A transmission line with electric oscillating circuits in the lower part: the depicted case is that of open-end termination. The two systems are described by the same dynamical equations.

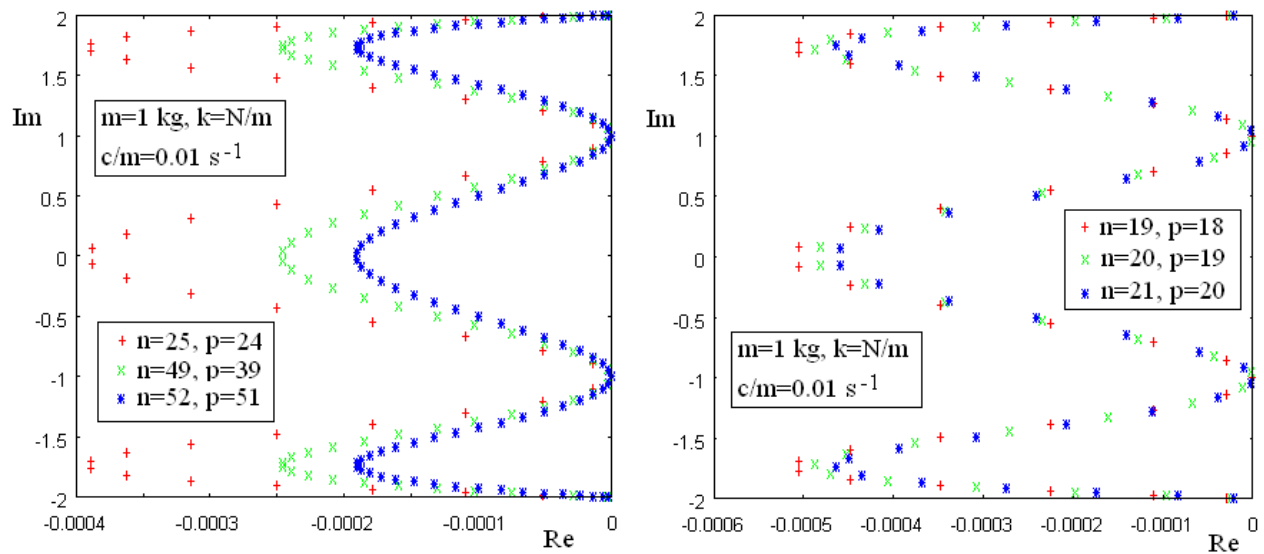

Fig.2: Behaviour of real (x-axis) and imaginary (y-axis) parts of eigenvalues, for corresponding choices of $n$ and $p$ as reported in the box, and for the specified physical parameters. The right part of the figure shows the behaviour of a chain ($n=19$ and $p=18$, red crosses) with pure imaginary eigenvalues, and those of two following chains $n=20$ and $n=21$, which have no undamped modes. Eigenvalues have a rather small real part, different from zero.

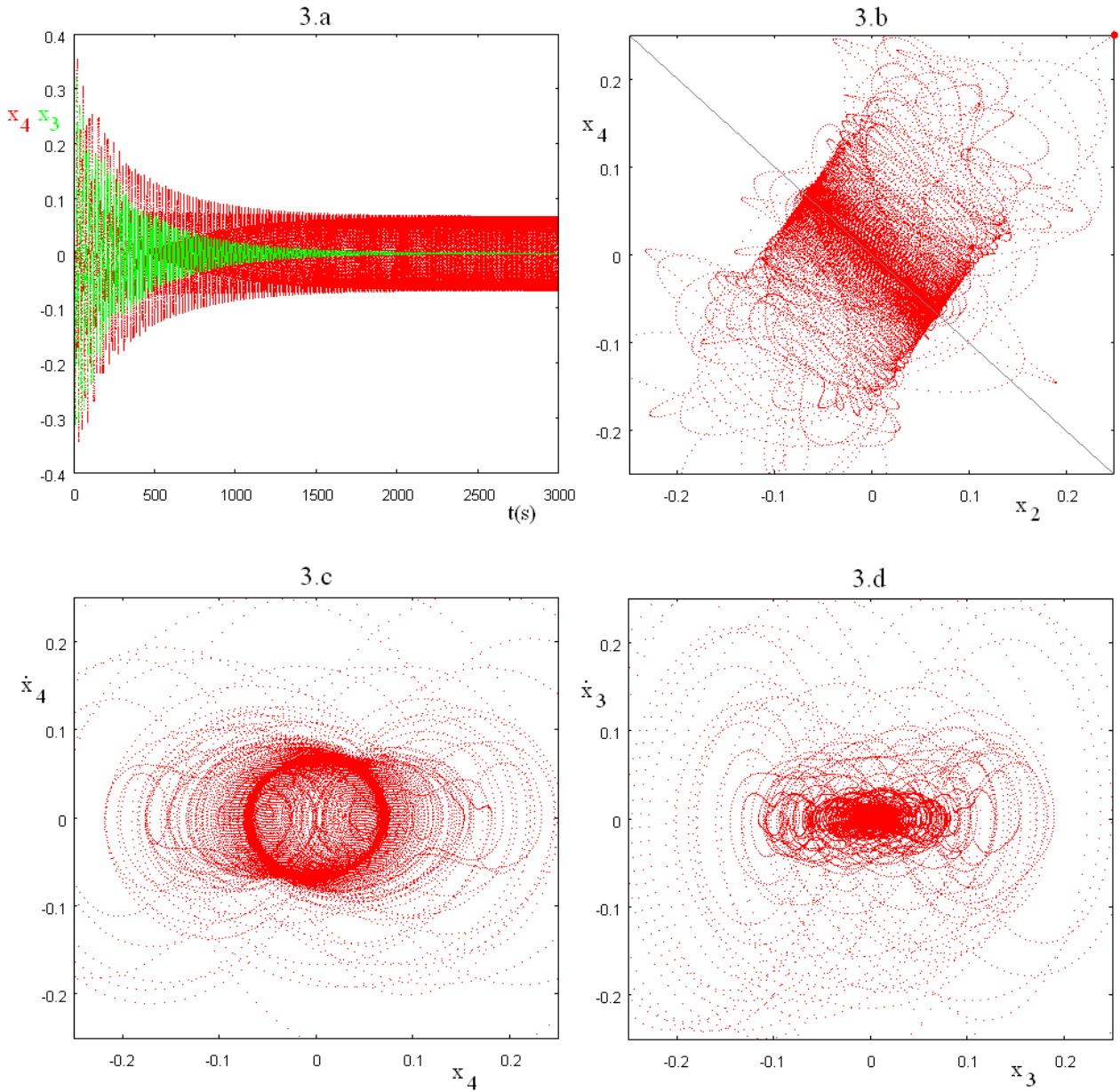

Fig.3: This image concerns the chain with 4 masses ($m_1, m_2, m_3, m_4$), the third ($m_3$) damped. Panel 3.a reports the behaviour of the third ($m_3$) and fourth masses ($m_4$) as a function of time, for $m$=1 kg, $k$=1 N/m, $c$=0.01 s$^{-1}$. $x_3$ is damped. 3.b reports the behaviour of the two positions ($x_2, x_4$) of oscillating masses ($m_2, m_4$), with time as parameter. The initial point of the motion is at the right upper corner marked with the red dot, whereas the motion limit for asymptotic times is coincident with central part of the grey diagonal line. 3.c and 3.d report the phase space diagrams for masses $m_4$ and $m_3$ respectively. Note the limit-cycle in 3.c. For 3.d, the limit is the central point.

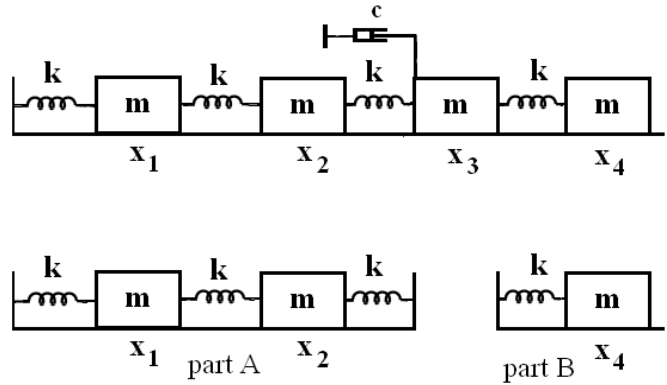

Fig.4: When only the undamped mode survives, the third mass becomes a node of the chain. This allows to view the chain as composed by two sub-chains (parts A and B). They oscillate synchronized at the same frequency and with a phase suitable to let mass 3 at its equilibrium position.

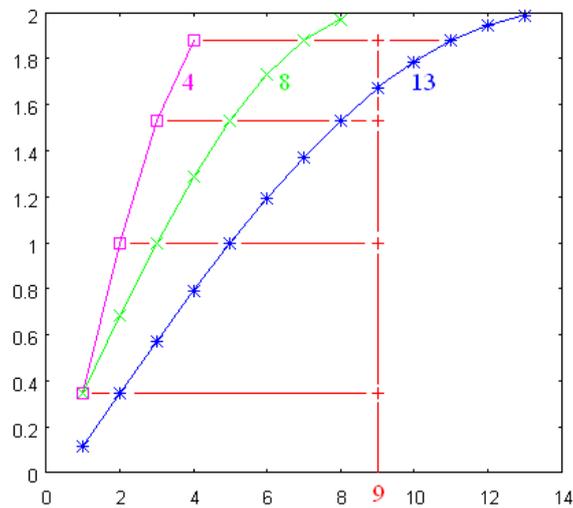

Fig.5: The same as in Table II, with the comparison of the dispersion of the chain without dampers with $n=13$, and the dispersion of the two sub-chains with $n=4$ and $n=8$. The red crosses are the modes appearing when the mass at position 9 is damped. The physical parameters are $m=1$ kg, $k=1$ N/m, $c=0.01$ s$^{-1}$.

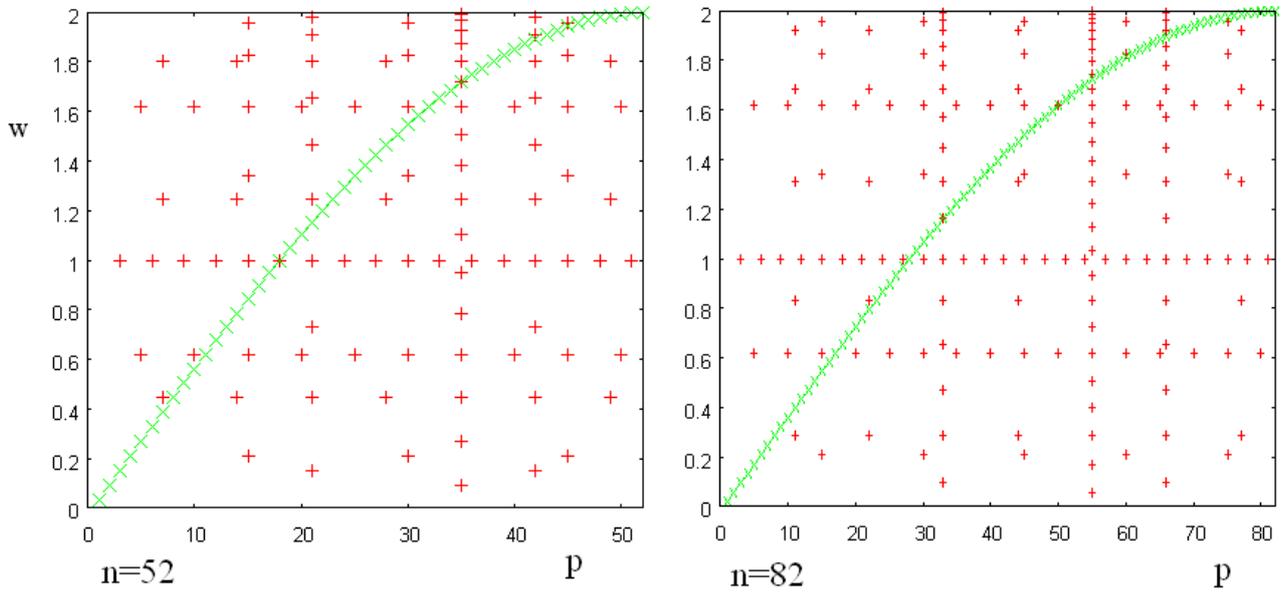

Fig.6: The red crosses are the values of the undamped modes appearing when the mass at position *p* is damped. The green points represent the dispersion of the chain without dampers. Physical parameters are $m=1$ kg, $k=1$ N/m, $c=0.01$ s$^{-1}$.